\documentclass{iau}

\usepackage{graphicx} 

\title[IAUS291.~~Magnetars: neutron stars with huge magnetic storms]{Magnetars: neutron stars with huge magnetic storms}
\author[N. Rea] %% short author list %%
{Nanda Rea}

\affiliation{Institut de Ci\`encies de l'Espai (CSIC--IEEC), Campus UAB, Facultat de Ci\`encies, \\Torre C5-parell, E-08193 Barcelona, Spain \\
email: {\tt  rea@ice.csic.es}
}

\pubyear{2012}
\volume{291}  %% insert here IAU Symposium No.
\jname{\mbox{Neutron Stars and Pulsars: Challenges and Opportunities after 80 years}}
\editors{J. van Leeuwen, ed.} 
\begin{document}

\maketitle

%% -- Abstract ----------------------------------
\begin{abstract}
Among the many different classes of stellar objects, neutron stars
provide a unique environment where we can test (at the same time) our
understanding of matter with extreme density, temperature, and
magnetic field. In particular, the properties of matter under the
influence of magnetic fields and the role of electromagnetism in
physical processes are key areas of research in physics. However,
despite decades of research, our limited knowledge on the physics of
strong magnetic fields is clear: we only need to note that the
strongest steady magnetic field achieved in terrestrial labs is some
millions of Gauss, only thousands of times stronger than a common
refrigerator magnet.  In this general context, I will review here the
state of the art of our research on the most magnetic objects in the
Universe, a small sample of neutron stars called magnetars. The study
of the large high-energy emission, and the flares from these strongly
magnetized ($\sim10^{15}$ Gauss) neutron stars is providing crucial
information about the physics involved at these extremes conditions,
and favoring us with many unexpected surprises.

%% add here a maximum of 10 keywords, to be taken form the file <Keywords.txt>
\end{abstract}

\firstsection % if your document starts with a section,
              % remove some space above using this command.
\section{Introduction}
\noindent
Neutron stars are the debris of the supernova explosion of massive stars, the existence of which was first theoretically predicted around 1930 (Chandrasekhar 1931; Baade \& Zwicky 1934) and then observed for the first time more than 30 years later (Hewish et al. 1968). They were predicted all along as very dense and degenerate stars holding about 1.4 solar masses in a sphere of 10km radius. We now know many different
flavors of these compact objects, and many open questions are still waiting for an answer after decades of studies. The neutron star population is dominated by radio pulsars (thousands of objects), however in the last decades several extreme and puzzling sub-classes of isolated neutron stars were discovered: Anomalous X-ray Pulsars (AXPs), Soft Gamma Repeaters (SGRs; see Mereghetti 2008), Rotating Radio
Transients (RRATs; Keane \& McLaughlin 2011), X-ray Dim Isolated Neutron stars (XDINSs; Turolla 2009), and Central Compact Objects (CCOs; Mereghetti 2011). The large amount of different acronyms might already show how diverse is the neutron star class, and on the other hand, how far we are from a unified scenario. These objects are amongst the most intriguing populations in modern high-energy astrophysics and in physics in general. They are precious places to test gravitational and particle physics, relativistic plasma theories, as well as strange quark states of matter and physics of atoms and molecules embedded in extremely high magnetic fields (impossible to be reproduced on Earth).
Since their discovery in the late sixties, about 2000 rotational powered pulsars are known to date, thanks to numerous surveys using single dish radio antennas around the world (Parkes, Green Bank, Jodrell Bank, Arecibo), with periods ranging from about 1.5 ms to 12 s (see Figure 4, and the ATNF on-line catalog: Manchester et al. 2005), and they have magnetic fields ranging between  $\sim 10^{10} - 10^{15}$\,Gauss.  The energy reservoir of all those pulsars is well established to be their rapid rotation, having a rotational luminosity  $L_{\rm rot} \sim 4\pi^2 I \dot{P} /P^{3} \sim 3.9\times10^{46} \dot{P} /P^3$ erg/s . A key ingredient to activate the radio emission is the acceleration of charged particles, which are extracted from the star's surface by an electrical voltage gap. The voltage gap forms due to the presence of a dipolar magnetic field co-rotating with the pulsar, and it is believed to extend up to an altitude of $\sim 10^4$ cm with a potential difference $> 10^{10}$ statvolts. Primary charges are accelerated by the electric field along the magnetic field lines to relativistic speeds and emit curvature radiation. Curvature photons are then converted into electron-positron pairs and this eventually leads to a pair cascade which is ultimately responsible for the coherent radio emission we observe from radio pulsars. Very energetic pulsars are also observed until the gamma-ray range, most probably in the form of synchrotron photons coming from the acceleration in the so-called 'outer-gap' of the pulsar magnetosphere (Goldreich \& Julian 1969; Ruderman \& Sutherland 1975). All isolated pulsar rotational periods are increasing in time. This spin down is quantified by the braking index, $n = \Omega \ddot{\Omega}/\dot{\Omega}^2$ (where $\Omega = 1/P$). With this definition, under the assumption of pure dipole braking, we would expect all pulsars having $n = 3$.

%%%%%%%%%%%%%%%%%%%%%%%%%
\begin{figure}[t]
\begin{minipage}[b]{1\linewidth}
\centering
\includegraphics[width=0.9\textwidth]{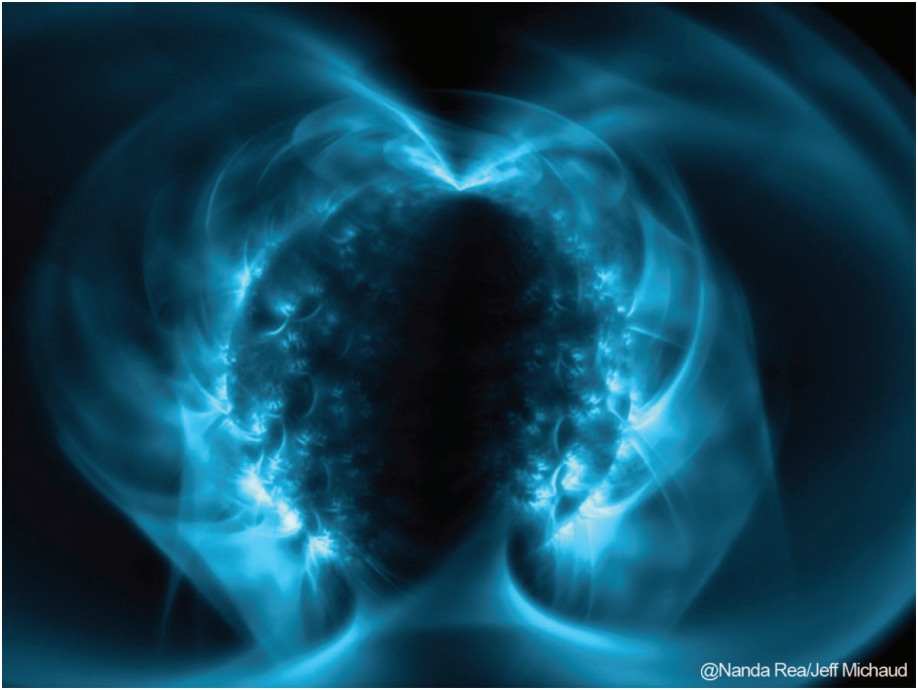}
\end{minipage}
\caption{Artistic image of the magnetic field of a magnetar close to its surface (Rea et al. 2012; published in Science as Editors' choice).}
\end{figure}
%%%%%%%%%%%%%%%%%%%%%%%%%

In this review we will report on the state of the art of the study of the strongest magnets in the Universe: the magnetars. However, before presenting these
ultra-magnetic objects, it is instructive to indicate how the magnetic field of isolated pulsars is commonly estimated. Assuming that pulsars slow down due to magnetic dipole radiation, the surface dipolar magnetic field ($B_{dip}$) can be estimated from the measured pulsar spin period P and its first derivative $\dot{P}$: $B_{dip}\sim  3.2\times10^{19} \sqrt{P\dot{P}}$ Gauss (where P is in units of seconds).
%%%%%%%%%%%%%%%%%%%%%%%%%
\begin{figure}[t]
\includegraphics[width=\textwidth]{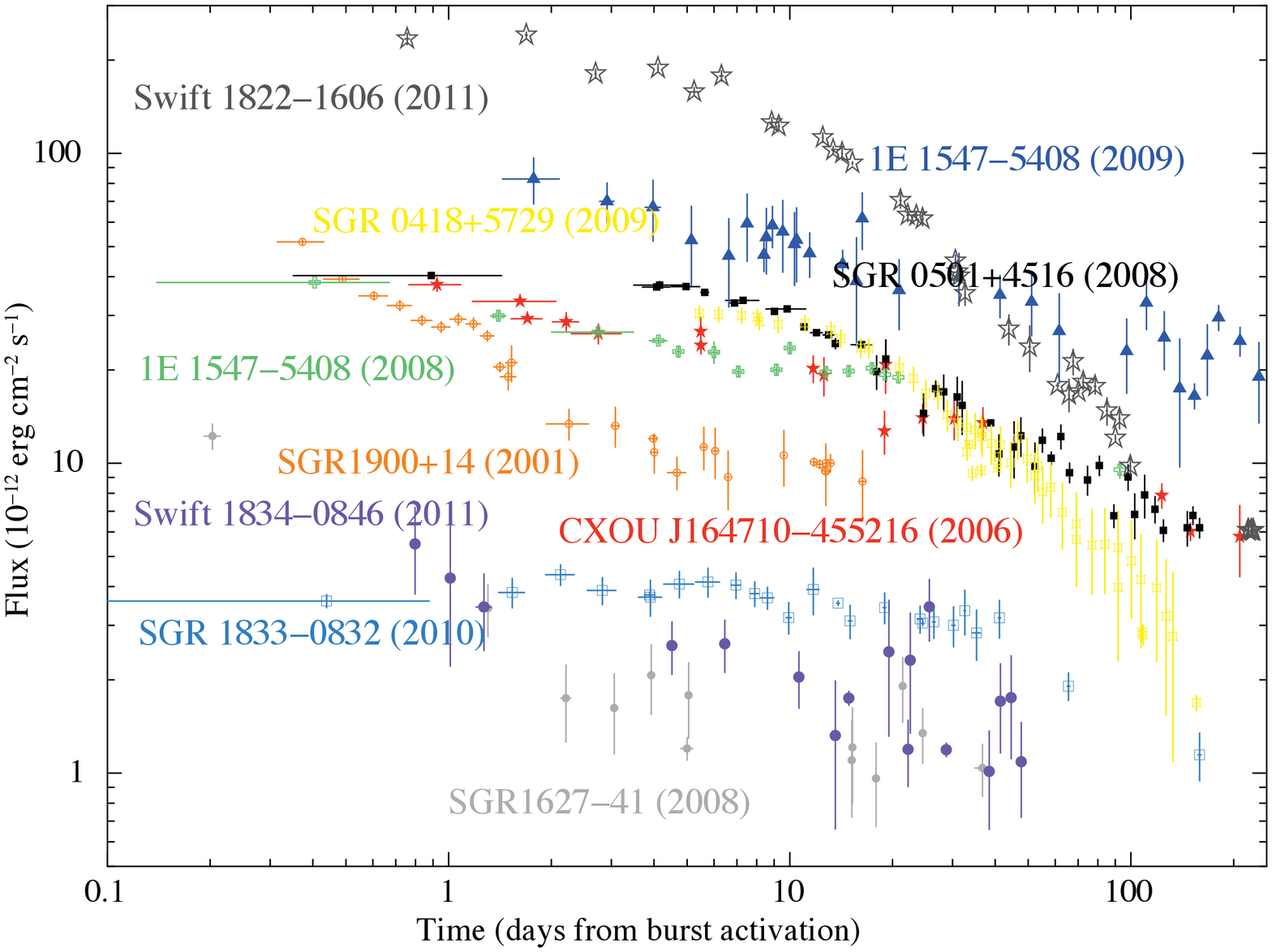}
\caption{Flux evolution over the first $\sim200$\,days of all magnetar outbursts (only if observed with imaging instruments, and for which this period span is well monitored). Fluxes are reported in the 1--10 keV energy range, and the reported times are calculated in days from the detection of the first burst in each source. See Rea \& Esposito (2011) for the reference for each reported outburst.}
\end{figure}
%%%%%%%%%%%%%%%%%%%%%%%%%
The 'magnetars' (comprising AXPs and SGRs; Mereghetti 2008) are a small group of X-ray pulsars (about twenty objects with spin periods between 2--12\,s) the emission of which is very hardly explained by any of the scenarios for the radio pulsar or the accreting X-ray binary populations. In fact, the very strong X-ray emission of these objects ($L_{x} \sim10^{35}$ erg/s) seemed too high and variable to be fed by the rotational energy alone (as in the radio pulsars), and no evidence for a companion star has been found so far in favor of any accretion process (as in the X-ray binary systems).
Moreover, roughly assuming them being magnetic dipole radiator, their inferred magnetic fields appear to be as high as $B_{dip} \sim 10^{14} -10^{15}$ Gauss. They are then higher than the electron critical magnetic field, $B_{Q}= m^{2}_e c^{3} /eh\sim 4.4\times10^{13}$ G at which an electron gyro-rotating around such magnetic field line gains a cyclotron energy equal to its rest mass. At fields higher than B$_{Q}$, QED effects such as vacuum polarization or photon splitting, can take place (see Harding \& Lai 2006).

%%%%%%%%%%%%%%%%%%%%%%%%%
\begin{figure}[t]
\begin{minipage}[b]{1\linewidth}
\centering
\hspace{-0.8cm}
\hbox{
\includegraphics[width=7.1cm,height=7cm]{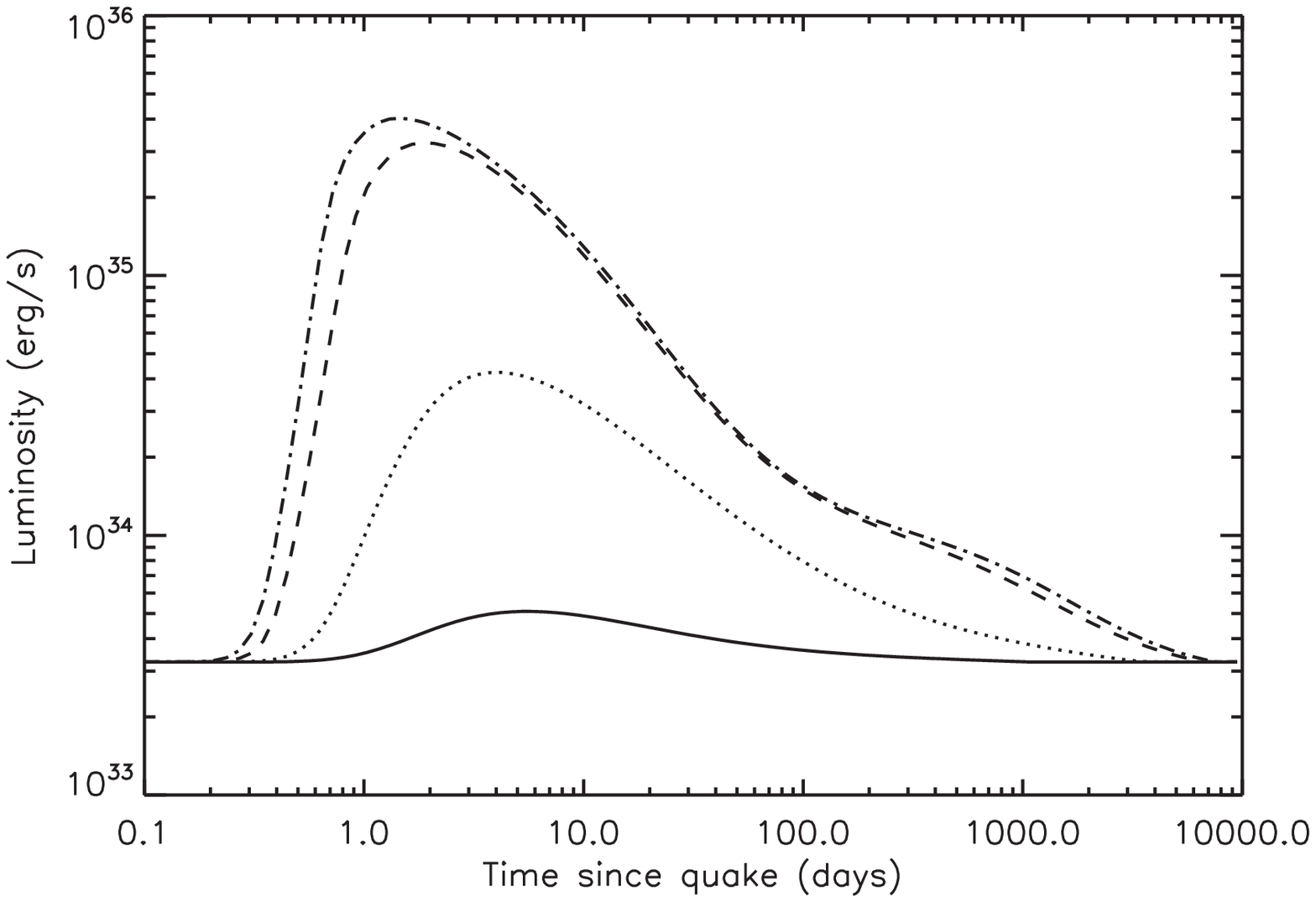}
%\hspace{0.3cm}
\includegraphics[width=6.9cm,height=7cm]{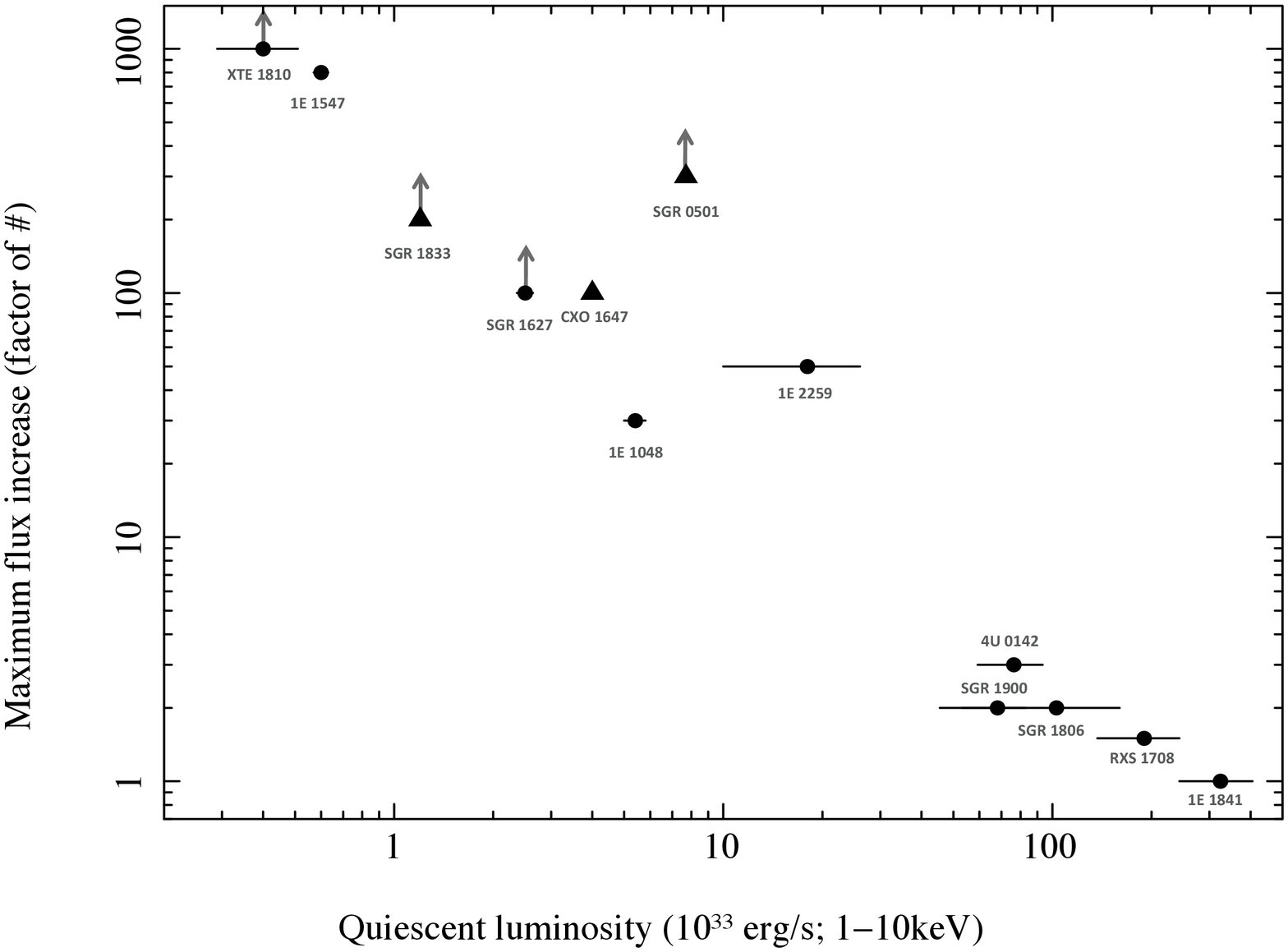}}
\end{minipage}
\caption{{\em Top panel}: Luminosity vs. time after energy injection. The models
correspond to $E_{\rm oc}=1.7\times10^{41}$\,erg, (solid line), $1.7\times10^{42}$\,erg (dotted line), 
 $1.7\times10^{43}$\,erg (dashed line), and $1.7\times10^{44}$\,erg (dash-dotted line). {\em Bottom panel}: quiescent luminosity vs. outburst maximum flux increase (all in the 1-10 keV band), for all magnetars showing bursts, glitches or outbursts.  See Pons \& Rea 2012 for further details.}
\end{figure}
%%%%%%%%%%%%%%%%%%%%%%%%%

Because of these high B fields, the emission of magnetars was thought
to be powered by the decay and the instability of their strong fields
(Duncan \& Thompson 1992; Thompson \& Duncan 1993, 1995). This
powerful X-ray output is usually well modeled by a thermal emission
from the neutron star hot surface (about $3\times10^6$ Kelvin)
reprocessed in a twisted magnetosphere through resonant cyclotron
scattering, a process favored only under these extreme magnetic
conditions (Thompson, Lyutikov \& Kulkarni 2002; Nobili, Turolla \&
Zane 2008; Rea et al. 2008; see Figure 1 for an artistic
representation of a magnetar). On top of their persistent X-ray
emission, magnetars emit very peculiar flares on short timescales
(from fraction to hundreds of seconds) emitting a large amount of
energy ($10^{40}-10^{46}$ erg; the most energetic Galactic events
after the supernova explosions).  They are probably caused by large
scale rearrangements of the surface/magnetospheric field, either
accompanied or triggered by fracturing of the neutron-star crust, as a
sort
of stellar quakes. Furthermore, magnetars also show  large outbursts
where their steady emission can be enhanced up to $\sim$1000 times its
quiescent level (see Figure 2, and see Rea \& Esposito 2010 for recent
review on transient magnetars). From the few well-monitored events, we
are starting to understand how those outbursts are produced.  They are
caused by similar crustal fractures as the shorter flares, accompanied
by strong surface heating, and often by the appearance of additional
hot spots on the neutron-star surface. This is what may cause large
spectral changes during outbursts, pulse profile variability, and
different cooling patterns  depending on the outburst. We have
recently started to model that outburst decay, and much important
physical information is slowly emerging, i.e. that all outbursts saturate at $\sim10^{36}$ erg/s, due to neutrino cooling processes, and regardless the source quiescent level. This discovery makes magnetar outbursts potential standard candles  (Pons \& Rea 2012; see Figure 3). 

%%%%%%%%%%%%%%%%%%%%%%%%%
\begin{figure}[t]
\begin{minipage}[b]{1\linewidth}
\centering
\includegraphics[width=1.0\textwidth]{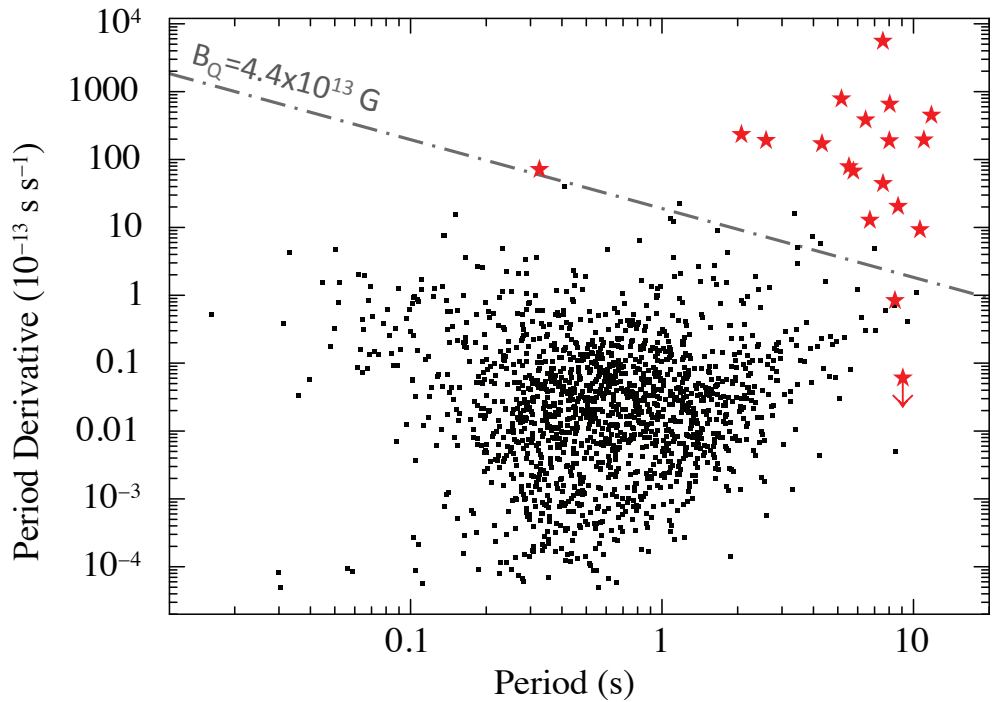}
\end{minipage}
\caption{$P$--$\dot{P}$ diagram for all known isolated pulsars.  Black squares represent normal radio pulsars, and red stars are all pulsars showing magnetar-like emission. The two newly discovered low-B magnetars: Swift J1822.3--1606 (Rea et al. 2012), and  SGR 0418+5729 (Rea et al. 2010) are also reported, as well as the electron quantum magnetic field (dash-dotted grey line).}
\end{figure}
%%%%%%%%%%%%%%%%%%%%%%%%%

\section{Hints for the connection between magnetars and radio pulsars}

\noindent
In the past few years, new discoveries started to shed light on a possible connection between magnetars and the typical radio pulsar population, weakening the strong distinction between these two classes, while pointing to a continuum of magnetar-like emission in the neutron star population. Below we list a few of those key discoveries.

\smallskip
%%%%%%%%%%%%%%%%%%%%%%%%%
\begin{figure}[t]
\includegraphics[trim=20 20 35 20, clip, width=1.0\textwidth]{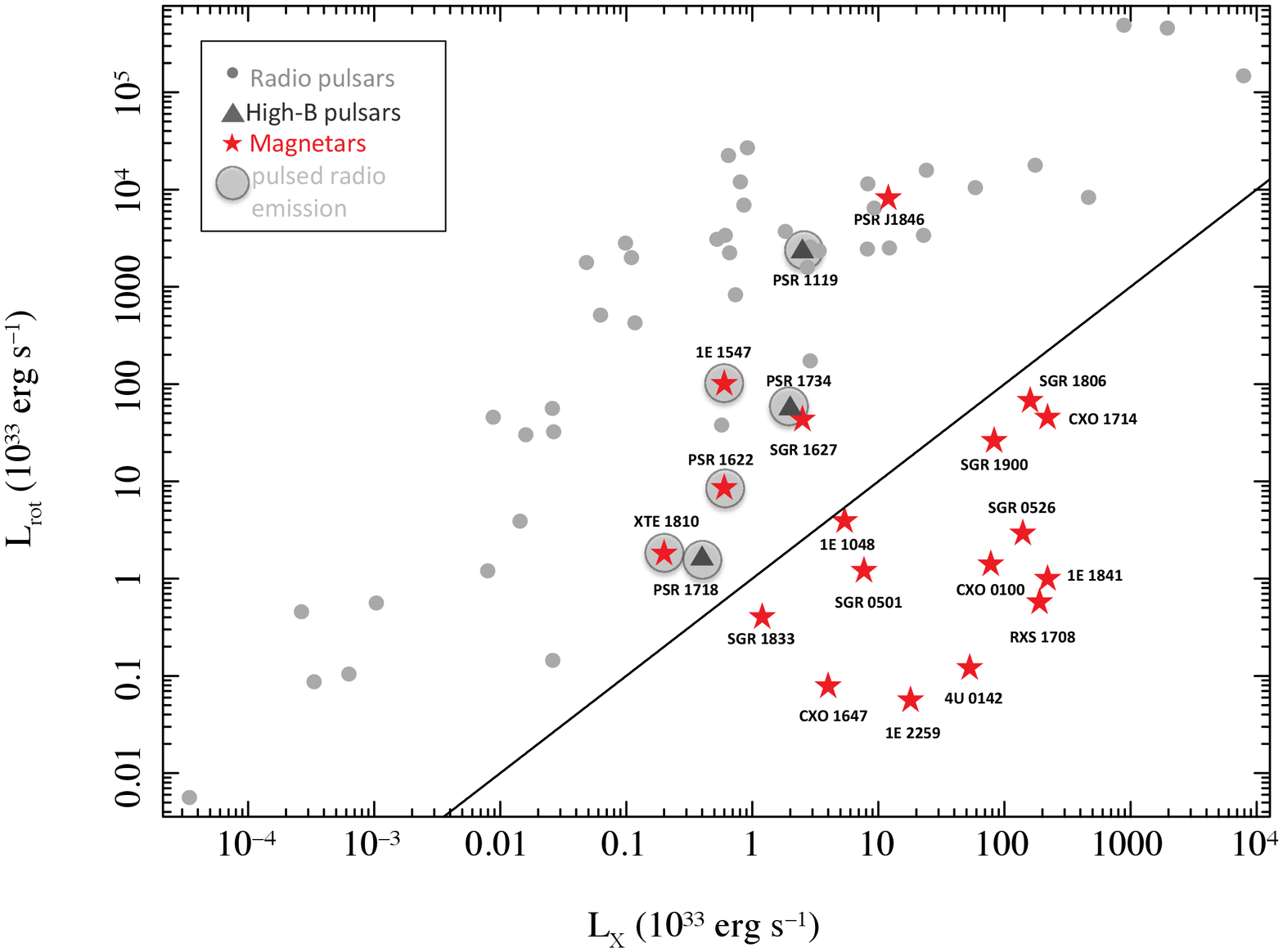}
\caption{X-ray luminosity versus the spin-down luminosity for all pulsars having a detected X-ray emission (grey filled circles), high-B pulsars (filled triangle), and the magnetars (red stars). Grey shaded circles mark the magnetars and high-B pulsars with detected pulsed radio emission, and the solid line shows $L_{x} = L_{rot}$ . X-ray luminosities are calculated in the 0.5--10 keV energy range, and for variable sources refer to the quiescent emission state.}
\end{figure}
%%%%%%%%%%%%%%%%%%%%%%%%%

\begin{itemize}

\item 
\noindent
Magnetars were believed to be radio-quiet sources for a few
decades. This was interpreted as the result of a photon splitting
process that is very efficient in magnetic fields stronger than the
critical electron field ($B_{Q}$; Baring 1998). The discovery in 2004 of transient magnetars coincided also with the discovery of radio pulsed emission from such sources (Camilo et al 2006; Levin et al. 2010). Magnetar pulsed radio emission, however, appeared to have different properties with respect to normal radio pulsars (flat radio spectra, large variability, connection with X-ray outbursts). This came as a big surprise, and started the idea of a possible connection between magnetars and the typical radio pulsars. Furthermore, recently a study of radio magnetars showed that despite the different characteristics, the radio emission can have the same physical mechanisms as for rotation-powered pulsars (Rea et al. 2012): powered by rotational energy (see also Figure 5), but with different observational properties possibly caused by a different path that a pair cascade might undertake when embedded in a mostly toroidal magnetic field.

\smallskip
\item 
Deep radio surveys discovered a few radio pulsars having dipolar fields larger than the $B_{Q}$ (see Figure 4). Although having magnetic fields in the magnetar range those objects were behaving as normal radio pulsars, and this was interpreted by a different magnetic geometry between the two classes. In 2008, bursting activity and an X-ray outburst were detected from a high-B pulsar, showing the presence of magnetar-like activity (Gavriil et al. 2008; Kumar \& Safi-Harb 2008). 

\smallskip
\item 
The extensive follow-up of transient magnetars undergoing an outburst
had allowed the most unexpected discoveries. In particular, prompted
by detection of typical magnetar-like bursts and a powerful outburst
of the persistent emission, a new transient magnetar was discovered in
2009, namely SGR 0418+5729. However, with great surprise after more
than a year of extensive monitoring, no period derivative was
detected, which led to an upper limit on the source surface dipolar
field of $B_{dip} < 7.5\times10^{12}$ Gauss (Rea et al. 2010). For the
first time we witnessed a magnetar with a low dipolar magnetic
field. This discovery demonstrated that not only a critical magnetic
field ($>B_{Q}$) was not necessary to have magnetar-like activity, but
many 'apparently' normal pulsars can turn out as magnetars at anytime (in fact the discovery of a second low-B magnetar soon followed; Rea et al. 2012; Scholz et al. 2012; see also Figure 4).

\smallskip
\item 
Advances in the measurement of pulsar breaking indexes showed the existence of objects with indexes $n$ smaller than 3, which would imply an increasing magnetic field with age under the common magnetic breaking picture. In particular this is the case of the high-B pulsar PSR 1734-33 (Espinoza et al. 2011), discovered to have $n=0.9$ . This discovery favors models for which the magnetic field is buried into the crust by accretion in the first supernova phases, and start re-emerging during the pulsar life-time (Vigan\'o \& Pons 2012). A similar conclusion has been reached with the discovery of low-B fields in CCOs, whose young age and hot surface temperature are instead pointing to a strong buried magnetic field, despite what measured by their period and period derivatives (Halpern et al. 2007).

\end{itemize}

\section{Conclusions}

The above discoveries, among others, re-focus the attention on a few
important ingredients of neutron star physics: i) the surface dipolar
magnetic field strength cannot be the only parameter driving their
magnetar or radio pulsar nature, ii) magnetars can behave as radio
pulsars and vice-versa, possibly powered by a similar mechanism
sustained by rotational energy, and iii) an internal strong magnetic
field is required to explain the low braking indexes of a few radio
pulsars, as well as the emission of the compact central objects,
despite the rather low dipolar magnetic field component.

These discoveries show that extremely strong magnetic fields may be
very common among the pulsar population, rather than an
exception. This might imply that supernova explosions should be
generally able to produce such strong magnetic fields, hence that most
massive stars are either producing fast rotating cores during the
explosion to activate the dynamo, or are strongly magnetized
themselves (i.e. 1\,kGauss at least). Furthermore, in this scenario
several gamma-ray bursts (not only an irrelevant fraction), might
indeed be due to the formation of magnetars, and the gravitational
wave background produced by magnetar formation should then be larger
than predicted so far (important for future instruments as
Advanced-LIGO).

\vspace{1.5cm}

NR is supported by a Ram\'on y Cajal research position to CSIC, and by the grants AYA2009- 07391, SGR2009-811, TW2010005 and iLINK 2011-0303, and she is indebted to Jos\'e Pons, Gian Luca Israel, Paolo Esposito, Roberto Turolla, Silvia Zane, Daniele Vigan\'o, Rosalba Perna, Diego Torres,  and to many others, for the always exciting and pleasant time spent together to understand the physics behind the bewildering magnetar emission. NR wishes also to thank Dick Manchester, all the speakers, and the IAU organizers for the excellent meeting.

\end{document}